\documentclass[12pt,pctex32]{article}
\usepackage{fractal,overcite}
\usepackage{amsmath,amsfonts}
\usepackage{graphicx}
\renewcommand{\thefootnote}{\fnsymbol{footnote}}

\begin{document}

\title{Characterization of Laser Propagation Through Turbulent Media by Quantifiers Based on the Wavelet Transform\protect\footnotemark}

\addtocounter{footnote}{-1}
\renewcommand{\thefootnote}{\emph\alph{footnote}}

\author{Luciano Zunino\protect\footnotemark, {Dar\'{\i}o G. P\'erez\protect\footnotemark} and Mario Garavaglia\protect\footnotemark}
\address{Departamento de F\'{\i}sica,
         Facultad de Ciencias Exactas,\\
                 Universidad Nacional de La Plata (UNLP) y Centro de Investigaciones \'Opticas, CC.  124 Correo Central, 1900 La Plata, Argentina.}

\author{Osvaldo A.  Rosso\protect\footnotemark}
\address{Instituto de C\'alculo,
                  Facultad de Ciencias Exactas y Naturales,\\
                  Universidad de Buenos Aires (UBA). \\
                  Pabell\'on II, Ciudad Universitaria. 
                  1428 Ciudad de Buenos Aires, Argentina. }

\renewcommand{\thefootnote}{\fnsymbol{footnote}}
\footnotetext[1]{This work was partially supported by Consejo Nacional de Investigaciones 
Cient\'{\i}ficas y T\'ecnicas (CONICET), Argentina. }
\renewcommand{\thefootnote}{\emph\alph{footnote}}
\footnotetext[1]{\textsf{lucianoz@ciop.unlp.edu.ar}}
\footnotetext[2]{\textsf{dariop@ciop.unlp.edu.ar}}
\footnotetext[3]{\textsf{garavagliam@ciop.unlp.edu.ar}}
\footnotetext[4]{\textsf{rosso@ic.fcen.uba.ar,rosso@ba.net}}

\begin{abstract}
The propagation of a laser beam through turbulent media is modeled as a fractional Brownian motion (fBm). Time series  corresponding to the center position of  the laser spot (coordinates $x$ and $y$) after traveling  across  air in turbulent  motion, with  different  strength, are  analyzed  by the    wavelet   theory.  Two quantifiers are calculated, the  Hurst exponent, $H$, and the mean Normalized Total Wavelet Entropy, $\tilde{S}_{\text{WT}}$ .  It  is  verified that  both  quantifiers give complementary information about the turbulence state.
\end{abstract}
\bigskip

\section{Introduction}
\label{sec:intro}

Wavelets-based  tools  have been shown to be well-suited to fractal processes and 
their analysis.  
This is mainly  due  to the  fact  that  the  wavelet transform  incorporates  in  
its definition  two basic features, time and scale, which  are  of primary  importance  
for  fractal processes.   
Another  remarks concerns  the  structure of the wavelet  transform  which,  by
construction,  builds  a  signal  by  successive  refinements, starting  from  a  
coarse approximation and adding  finer  and finer  details  at each step.  
Such a procedure  is  of  course reminiscent of basic fractal constructions, thus 
suggesting we should make use of wavelets to synthesize a fractal process 
\cite{Aubert98,Degaudenzi00,Frantziskonis00}. 
 
Since  the earlier leading work of Mandelbrot\cite{Mandelbrot74,Mandelbrot75} there  
is  a lot  of  evidence  that  several  facets  of  fully  developed turbulent  flows  
are  fractals.  
Most of the  applications  of fractals  to  turbulence have been devoted  to  the  
study  of subsets of the region occupied by turbulent flow where a given property   
is   satisfied.   
For  example,  the  turbulent/non-turbulent interface, some constant property surfaces 
(such  as the  iso-velocity  and  iso-concentration  surfaces)  and  the structure  of 
spatial distribution of dissipation  rates  have been  characterized  as  fractals  or  
multifractals  and  the dimension  has been measured 
\cite{Sreenivasan86,Meneveau91,Scotti95}.  
Then, it is  natural  to apply wavelet tools for analyzing atmospheric turbulence
\cite{Hudgins93,Qian96,Papanicolaou02,Katul01}. 

The  fractional Brownian motion (fBm) was introduced by  Mandelbrot and  Van  Ness 
\cite{Mandelbrot68} as an example of a process which contains  an infinite domain of 
dependence with the intention of explaining the  results  reported by Hurst in 1951
\cite{Hurst51}.  
fBm as a model for turbulence  is not  new\cite{Papanicolaou02,Katul01,Perez03}.  
In particular, the  fBm can be introduced as an alternative  model for the turbulent  
index  of  refraction,  and it can be shown  that these processes allow to reconstruct 
most of the index properties\cite{Perez03}. 

This  work reports on the basic characteristics of  the  centroid position of a laser spot 
after the light has traveled through turbulent media. It is analyzed as a stochastic process, within  a mathematically defined theoretical  model: the fractional Brownian  motion. 

That is, given a component of the centroid position $f$ as a time $t$ function, the geometrical nature of the graph $(t,f(t))$ is studied.  Afterwards, the wavelet theory is used to characterize this centroid. Two quantifiers are obtained: the Hurst exponent, $H$, and the mean Normalized Total Wavelet Entropy, $\tilde{S}_{\text{WT}}$. Their behaviors are compared; the analysis shows they describe different properties of the turbulence.

\section{Fractional Brownian motion}
\label{sec:fBm}

The fractional Brownian motion  of exponent  $H$ (Hurst exponent), with $0<H<1$,
is a zero-mean Gaussian process $B_H(x)$ with $x\in \mathbb{R}$ such that:
\renewcommand{\theenumi}{\emph{\textbf{\alph{enumi}}}}
\begin{enumerate}
\item $B_H(0)=0$. 
\item  The difference $B_H(x+\Delta x) - B_H(x)$ has finite dimensional  normal distribution. 
\end{enumerate}
This means that their increments are stationary Gaussian processes whose variance is
\begin{equation}
\mathbb{E}\left[(B_H(x+\Delta x)-B_H(x))^2\right]=\sigma^2\left|\Delta x\right|^{2H},
\label{eq:fBmvariance}
\end{equation}
where $\sigma$ is a parameter---$\mathbb{E}[\:\cdot\:]$ is, of course, the average. 
The nonstationary character of the fBm is evidenced by its covariance function
given by 
\begin{equation}
\mathbb{E}\left[B_H(x)B_H(y)\right]=
{\frac{\sigma^2}{2}} \left\{\left|x\right|^{2H}+\left|y\right|^{2H}-\left|x-y\right|^{2H}\right\}. 
\label{eq-1}
\end{equation}
The covariance of future increments with past ones is:
\begin{eqnarray}
 \rho_H\left(\Delta x\right) &=& \mathbb{E}\left[(B_H(x)-B_H(x-\Delta x))\cdot\left(B_H(x+\Delta x)-B_H(x)\right)\right] 
                 \nonumber \\
                  &=&{\sigma^2}(2^{2H-1}-1)\left|\Delta x\right|^{2H}. 
\label{eq-3}
\end{eqnarray}
Note that $ \rho_H$ is independet of $x$ and the parameter $H$ determine the correlation
of the increments. 

For $H=1/2$ the correlation of past and future increments vanishes for all $x$, as is 
required for an independent random process.  
However, for $H \not = 1/2$ one has $\rho_H(\Delta x) \not = 0$.   
For $H > 1/2$ this quantity is positive and the process  is called persistent\cite{Feder88}.  
In this case, an increasing trend in  the past implies on the average an  increasing  trend  in  the   
future and, conversely,  a  decreasing trend in the past  implies  on  the average a 
continued decrease in the future.  For $H < 1/2$ the process is called antipersistent.  
Now, an increasing trend  in  the past implies on the average a decreasing trend in the future, while a decreasing trend in the past makes on the average an increasing trend in the future---see Fig. {\ref{figura0}}. 

K.   Helland and Van Atta\cite{Helland78} were the first to study  the  Hurst exponent  
in grid generated turbulence by applying a rescaled-range  analysis---an usual measure of 
long-term persistence  in geophysical  time  series---to turbulence velocity records measured within a wind tunel.  They showed that  there  are  some deviations from $H = 1/2$---referred as the  \textit{Hurst phenomenon}. 

As  a  nonstationary process, the fBm does not have a spectrum defined in the usual sense; 
however, it is possible to  define an empirical power spectrum of the form:
\begin{equation}
S_{B_H}(f)={\frac{\sigma^2}{\left|f\right|^{2H+1}}}. 
\label{eq-4}
\end{equation}
This equation is not a valid power spectrum in the theory of stationary processes  since
it is a nonintegrable function, but it could be considered  as a generalized spectrum. 
Through this interpretation of Eq.  (\ref{eq-4}) a self-similarity relation can be shown for
the fBm.  That is for $B_H(x)$ with $H$ and $\sigma$ parameters, one has that $a^H B_H(x/a - b)$ have the  same  finite dimensional distributions for  all $a>0$ and  $b$.  The fractal dimension of sample functions of these processes is $D=2-H$.  

\section{Time-Frequency Analysis}
\label{sec:time-frequency}

\subsection{Wavelet transform}
\label{sec:Wavelet-Transform}

First introduced by Dennis Gabor \cite{paper:gabor}, wavelet analysis is a method which relies on the introduction of an
appropriate basis and a characterization of the signal by the distribution
of amplitude in this basis. 
If the basis is required to be a proper orthogonal basis, any arbitrary
function can be uniquely decomposed and the decomposition can be inverted
\cite{Daubechies92,Aldroubi96,Mallat99,Samar99}. 
Wavelet analysis is a suitable tool for detecting and characterizing specific
phenomena in time and frequency planes.

The \textit{wavelet} is a simple and quickly vanishing (compactly supported) oscillating function. Unlike sine and cosine of Fourier analysis, which are precisely localized in frequency but extend infinitely in time, wavelets are relatively localized in both time and frequency. Furthemore, wavelets are band-limited; they are composed of not one but a relatively limited range of several frequencies.

A \textit{wavelet family} $\psi_{a,b}$ is the set of elementary functions generated by dilations and translations of a unique admissible
 \textit{mother wavelet} $\psi (t)$:
\begin{equation}
\label{eq:wav1}
\psi _{a,b}( t )=\left|a\right|^{-1/2}\psi \left( \frac{t-b}{a} \right),
\end{equation}
where $a, b \in \mathbb{R}$, $a \neq 0$ are the scale and translation
parameters respectively, and $t$ is the time. 
As $a$ increases, the wavelet becomes narrower. 
Thus, one have a unique analytic pattern and its replications at different
scales and with variable time localization. 

The \textit{continuous wavelet transform} (CWT) of a signal
$\mathcal{S}(t) \in L^2(\mathbb{R})$ (the space of real square summable
functions) is defined as the correlation between the function
$\mathcal{S}(t)$ with the family wavelet $\psi_{a,b}$ for each $a$ and $b$:
\begin{equation}
\label{eq:wav2}
\left(W_{\psi}\mathcal{S}\right)(a,b) = \left|a\right|^{-1/2}
                         \int_{-\infty}^{\infty}\mathcal{S}(t)
                        {{\psi^*}\left(  \frac{t-b}{a} \right) }dt
                    = \langle\mathcal{S},\psi_{a,b}\rangle. 
\end{equation}
For special election of the mother wavelet function $\psi(t)$ and for the
discrete set of parameters,
$a_{j} = 2^{-j}$ and $b_{j,k} = 2^{-j} k$,  with $j, k \in \mathbb{Z}$
(the set of integers) the family
\begin{equation}
\label{eq:wav3}
\psi_{j,k}( t )=2^{j/2}\psi(2^jt-k)
\quad \quad \quad j,k\in\mathbb{Z},
\end{equation}
constitutes an orthonormal basis of the Hilbert space $L^2(\mathbb{R})$
consisting of finite-energy signals. 

\begin{figure} 
\begin{center}
\includegraphics[width=0.8\textwidth]{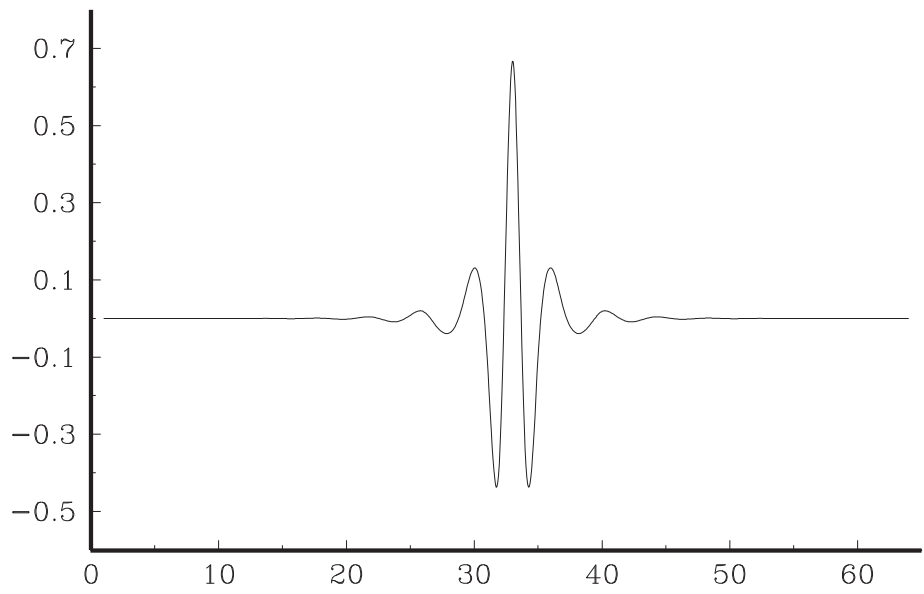}
\end{center}
\caption{Cubic spline wavelet. }
\label{figura0a}
\end{figure}

The correlated  \textit{decimated discrete wavelet transform}
(DWT) provides a nonredundant representation of the signal, and the values 
$\langle\mathcal{S},\psi_{a,b}\rangle$ constitute the coefficients in a
wavelet series. 
These wavelet coefficients provide relevant information in a simple way and
a direct estimation of local energies at the different scales. 
Moreover, the information can be organized in a hierarchical scheme of nested
subspaces called multiresolution analysis in $L^2(\mathbb{R})$. 
In the present work, we employ orthogonal cubic spline functions as mother
wavelets. 
Among several alternatives, cubic spline functions are symmetric and combine
in a suitable proportion smoothness with numerical advantages. 
They have become a recommendable tool for representing natural signals
\cite{Thevenaz00,Unser99}---figure \ref{figura0a} shows the cubic spline wavelet function used in this work. 

In what follows, the signal is assumed to be given by the sampled values $\mathcal{S} = \{x(n), n = 1,\cdots, M \}$, corresponding to an uniform time grid with sampling time $T_s$. For simplicity, the sampling rate is taken as $T_s = 1$. If the decomposition is carried out over all resolutions levels the wavelet expansion is:
\begin{equation}
\label{eq:wav4}
\mathcal{S}( t ) = \sum_{j= -N}^{-1}\sum_kC_j(k)\psi_{j,k}(t)
                  = \sum_{j= -N}^{-1}r_j(t)   \  ,
\end{equation}
where $N = \log_2 M$ and the wavelet coefficients $C_j(k)$ can be interpreted as the local
residual errors between successive signal approximations at scales $j$ and
$j+1$, and $r_{j}(t)$ is the \textit{residual signal} at scale $j$. 
It contains the information of the signal $\mathcal{S}(t)$ corresponding
to the frequencies $2^{j-1} \omega_s \leq | \omega | \leq 2^j \omega_s$ with $\omega_s$ the sampling frequency. 

\subsection{Relative Wavelet Energy}
\label{sec:RWE}

Since the family $\{ \psi_{j,k}(t) \}$ is an  \textit{orthonormal} basis for
$L^2(\mathbb{R})$, the concept of energy is linked with the usual notions
derived from  Fourier's theory.  
The wavelet coefficients are given by 
$C_j(k) = \langle \mathcal{S}, \psi_{j,k} \rangle$ and the corresponding asociated
energy will be its square.  
The energy at each resolution $j= -1, \cdots, -N$, will be 
\begin{equation}
\mathcal{E}_j={\frac{1}{N_j}}\sum_kC^2_j(k),
\label{eq:wav6} 
\end{equation}
where $N_j$ represents the number of wavelet coefficients at resolution $j$. 
The total energy can be obtained in the fashion
\begin{equation}
\label{eq:wav7}
\mathcal{E}_{ \textit{tot}}=\sum_{j<0}\mathcal{E}_j. 
\end{equation}
Finally, we define the normalized $p_j$ values, which represent the
 \textit{Relative Wavelet Energy} (RWE) by
\begin{equation}
\label{eq:wav8}
p_j=\mathcal{E}_j/\mathcal{E}_{ \textit{tot}},
\end{equation}
for the resolution levels $j = -1, -2, \cdots ,-N$. 
The $p_j$ yield, at different scales, the probability distribution for the
energy. 
Clearly, $\sum_{j} p_j = 1$ and the distribution $\{ p_j \}$ can be
considered as a time-scale density that constitutes a suitable tool for
detecting and characterizing specific phenomena in both the time and the
frequency planes. 

\subsection{Wavelet entropy}
\label{sec:TWS}

The Shannon entropy\cite{Shannon48} gives a useful criterion for analyzing
and comparing probability distribution. 
It provides a measure of the information contained in any distribution. 
We define the  \textit{Normalized Total Wavelet Entropy} (NTWS) 
\cite{PRE3,Method01,Mairal02} as
\begin{equation}
\label{eq:wav9}
  S_{WT}=  -\sum_{j<0}  p_j  \cdot \ln p_j/ S^ \textit{max},
\end{equation}
where
\begin{equation}
\label{eq:wav10}
  S^ \textit{max}=\ln N. 
\end{equation}

The NTWS appears to be a measure of the degree of order-disorder of the signal. It provides useful information about the underlying dynamical process associated with the signal\cite{Mairal02}. Indeed, a very ordered process can be represented by a periodic mono-frequency signal (signal with a narrow band spectrum). A wavelet representation of such a signal will be resolved at one unique wavelet resolution level, i. e., all RWE will be (almost) zero except at the wavelet resolution level which includes the representative signal frequency. For this special level the RWE will be (in our chosen energy units) almost equal to one. As a consequence, the NTWS will acquire a very small, vanishing value. A signal generated by a totally random process or chaotic one can be taken as representative of a very disordered behavior. This kind of signal will have a wavelet representation with significant contributions coming from all frequency bands. Moreover, one could expect that all contributions will be of the same order. Consequently, the RWE will be almost equal at all resolutions levels, and the NTWS will acquire its maximum possible value. 

\begin{figure} 
\begin{center}
\includegraphics[width=0.8\textwidth]{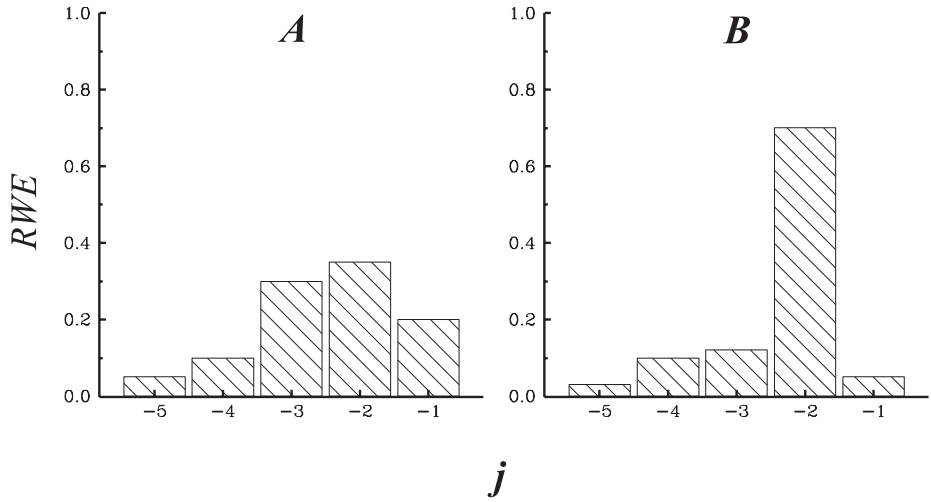}
\end{center}
\caption{Relative wavelet energy (probability) distributions corresponding to five wavelet resolution levels ($j=-5,\dots, -1$). Distribution $A$, $\{p_j\}=\{0.05,0.10,0.30,0.35,0.20\}$; distribution $B$, $\{p_j\}=\{0.03,0.10,0.12,0.70,0.05\}$. The NTWS values for this distribution are NTWS$(A)=0.888$  and NTWS$(B)=0.614$.}
\label{figura0b}
\end{figure}

Figure \ref{figura0b} presents two different relative wavelet energy (probability) distribution corresponding to five wavelet resolution levels  ($j=-5,\dots, -1$). It is clear from the figure that distribution $A$ presents broad band spectrum. In contrast, distribution $B$ shows a clear dominance of the resolution level $j=-2$. According to the description given above, for the NTWS the following relations can be expected: NTWS$(A) >$ NTWS$(B)$. This is observed in the numerical values given at the figure. 

\subsection{Wavelet quantifiers time evolution}
\label{sec:Time-Evolution}

In order to follow the temporal evolution of the quantifiers defined above, RWE and NTWS, the analyzed signal is divided into $i$ non-overlapping temporal windows  with length $L$ ($i = 1, \cdots, N_T$, with $N_T = M / L$).  Afterwards, appropriate signal-values for these quantifiers are assigned to the middle point of each time window. In the case of a diadic wavelet decomposition, the number of wavelet coefficients at resolution level $j$ is two times smaller than at the previous, $j+1$, one. The minimum length of the temporal window $L$ will therefore include at least one wavelet coefficient at each level. 

The wavelet energy at resolution level $j$ for the time window $i$ is given
by
\begin{equation}
\label{eq:wav14}
\mathcal{E}^{(i)}_j={\frac{1}{N_j}}
\sum_{k=(i-1)\cdot L + 1}^{i \cdot L}C_j^2(k) 
\qquad
\mbox{with $i = 1, \cdots , N_T$} \ ,
\end{equation}
where $N_j$ represents the number of wavelet coefficients at resolution level $j$ 
corresponding to the time window $i$;
while the total energy in this time window will be
\begin{equation}
\label{eq:wav15}
{\mathcal E}^{(i)}_{ \textit{tot}} =\sum_{j < 0} \mathcal{E}^{(i)}_j \ . 
\end{equation}
The time evolution of RWE and NTWS  will be given by:
\begin{equation}
\label{eq:wav16}
p^{(i)}_j=\mathcal{E}^{(i)}_j / \mathcal{E}^{(i)}_{ \textit{tot}},
\end{equation}
\begin{equation}
\label{eq:wav17}
 S_{WT}{(i)}=  -\sum_{j<0}  p^{(i)}_j \cdot \ln p^{(i)}_j/S^ \textit{max}. 
\end{equation}

In order to obtain a quantifier for the whole time period under analysis
\cite{Method01} the temporal average is evaluated. 
The temporal average of NTWS is given by
\begin{equation}
\label{eq:16}
\langle S_{ \textit{WT}}\rangle = \frac{1}{N_T}\sum_{i=1}^{N_T} S_{WT}^{(i)},
\end{equation}
and for the wavelet energy at resolution level $j$
\begin{equation}
\label{eq:17}
\langle \mathcal{E}_j\rangle = \frac{1}{N_T}\sum_{i=1}^{N_T}\mathcal{E}^{(i)}_j;
\end{equation}
then the total wavelet energy temporal average is defined as
\begin{equation}
\label{eq:18}
\langle \mathcal{E}_{ \textit{tot}} \rangle=\sum_{j < 0} \langle \mathcal{E}_j\rangle. 
\end{equation}
In consequence, a mean probability distribution $\{ q_j \}$
representative for the whole time interval (the complete signal) can be defined as
\begin{equation}
\label{eq:19}
q_j = \langle \mathcal{E}_j\rangle / \langle \mathcal{E}_{ \textit{tot}} \rangle,
\end{equation}
with $\sum_{j} q_j = 1 $ and the corresponding  mean NTWS as
\begin{equation}
\label{eq:20}
{\widetilde  S_{WT}}=-\sum_{j<0}  q_j \cdot \ln q_j/S^ \textit{max}. 
\end{equation}

\section{Fractional Brownian motion and Wavelet Transform}
\label{sec:fBmwavelet}

A relevant property of the wavelet based multiresolution analysis is the stationary character of the wavelet coeficient series corresponding to each level $j$ of resolution\cite{Datellis01}. Another important property is that the reconstruction of the original time series from the stationary series of wavelet coefficients reproduces the original signal with small error. 

In  relation with fractional Brownian motion it can  be  shown
that:
\begin{enumerate}
\item fBm is nonstationary but the wavelet coefficients  are stationary at 
each scale\cite{Flandrin89,Flandrin92};
\item fBm  exhibits positive long-range correlation in the range $1/2 < H < 1$ 
but wavelet coefficients have  a  correlation which can be arbitrarily reduced;
\item the self-similarity of fBm is directly reflected in its wavelet coefficients, 
whose variance varies as a  power law as a function of scale $j$\cite{Flandrin89,Flandrin92}
\begin{equation}
 \log_2 \left\{\mathbb{E}\left[C^2_j(k)\mid_{B_H}\right]\right\} \propto-(2H+1)j. 
 \label{eq2:1}
\end{equation}
\end{enumerate}

All  the  above properties evidence that wavelet  analysis  is naturally 
well-suited to fBm.  
Each of them provides in fact a key ingredient for a problem  of  major importance  
when analyzing fBm: the estimation of the Hurst exponent or  of  the related 
spectral exponent $\alpha = 2H+1$ \cite{Frantziskonis00,Datellis01}.  
Starting from  the above  mentioned observations that, in the wavelet  transform,
variance progression follows a power law across scales
\begin{equation}
\label{eq2:2}
 V\left[j\right]=\mathbb{E}\left[C^2_j(k)\mid_{B_H}\right]  \sim2^{-\alpha j},
\end{equation}
one  can  simply make use of the empirical variance estimators
at scale  $j$ (based on $N_j= 2^{-j} M$ coefficients for a sample of total length $M$)
\begin{equation}
\hat{V}\left[j\right]=\frac{1}{N_j}\sum_{k=1}^{N_j} C^2_j(k)\mid_{B_H}=\mathcal{E}_j \mid_{B_H},
\label{eq2:3}
\end{equation}
This is made possible because the fBm, when decomposed via the wavelet transform, becomes 
stationary at each scale. 

In this way an estimator of the parameter $H$ can be obtained by:  \textit{a)} estimating the variance of the wavelet coefficients with Eq. (\ref{eq2:3});  \textit{b)} plotting $\log_2\{{\hat V}[j]\}$ versus $j$ and fitting a minimum square line. The slope of the line give the estimator of $H$. 

Wavelet-based  estimators dedicated to fBm can  be  viewed  as versatile generalizations 
of previous techniques.  An important feature of fBm is that its increments are stationary and  such that  the structure function is proportional to $\left|\Delta x\right|^{2H}$---see Eq. (\ref{eq:fBmvariance})---thus  suggesting  that this variance should  be  estimated  in order to find $H$.  
Although feasible, this approach is faced with a difficulty due to long-range  dependence.    Classical (empirical) variance estimators are obviously poor  estimators in such a  
context  and  specific  estimators  have  to be designed\cite{Beran94,Taqqu95}.

\section{Experimental Setup and Data Adquisition}
\label{sec:experimental}

Time  series  corresponding to the fluctuations in the position of a laser  
beam's spot (wandering) over a screen, after propagation through laboratory generated 
convective turbulence, were recorded  with a  position sensitive detector located as screen 
at the end of the path. 
Twenty  records of  5000 spot beam coordinates measurements every five minutes were obtained.  
Temperature along the laser beam path, for each record,  were also measured and stored. 

The experimental measures were performed in the laboratory by producing convective turbulence 
over a path of lenght $L=1. 5$m with two electric heaters in a row and covering the path. 
We use a $10$mW continuous wave He-Ne laser (Melles Griot Model 05-LHP-991), wavelenght laser 
beam $\lambda = 632. 8$nm (red), with pointing stability, after 15 minutes of warmup, less than 
$30\mu$m and a beam divergence of $1. 24 \pm 5\%$mrad. 

\begin{figure} 
\begin{center}
\includegraphics[width=0.9\textwidth]{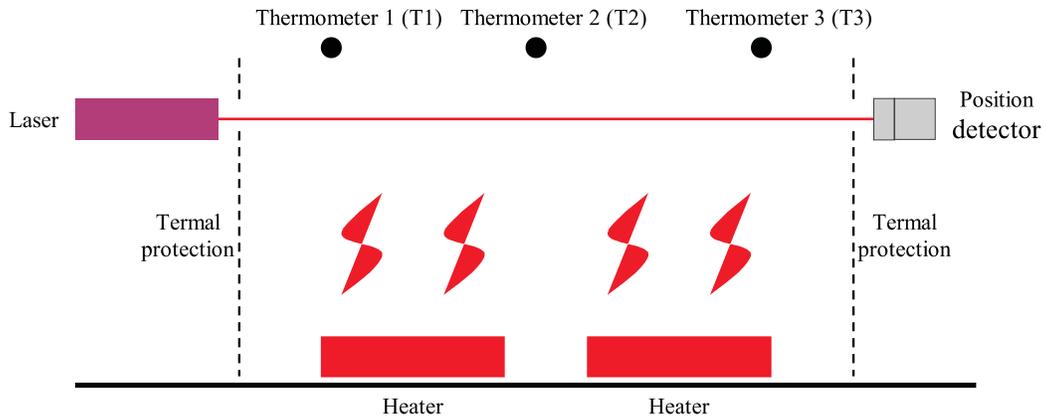}
\end{center}
\caption{Experimental setup.}
\label{figura1}
\end{figure}

Each electric heater was $50$cm long and $800$W of power.  
The height of the laser beam's propagation path was  $1$m above the electrical heaters and a
box of expanded polyurethane was used as a thermal protection for the equipment. 
Three  thermometers were allocated on the top of the box, $10$cm above the laser beam.  
They were used to determine the  temperature  in three  positions along the propagation path,  
see  Fig. \ref{figura1}.   
The geometrical experimental arrangement was similar to that used by Consortini et al. 
\cite{Consortini90} and, Consortini and O'Donnell\cite{Consortini91}. 

The  position sensor has a relative accuracy of the  order  of $2.5\mu$m so that very 
small position fluctuation can be measured. It was interfaced to a  computer which allowed to measure at a rate of about $800$ samples per second ($M=5000$ laser  spot coordinates  in approximately $7$s). Thus, with these coordinates stored on a hard disk the Hurst exponent and  
the mean NTWS were computed off line.

Three  different intensity  levels of convective turbulence were generated by changing  
the amount of heat dissipated for  each  electrical heater:

\begin{enumerate}
\item Normal turbulence, the electrical heaters were off;
\item Soft turbulence,  each electrical heaters  dissipated half of its available power;
\item Hard turbulence, each electrical heaters dissipated at the maximum available power. 
\end{enumerate}

\begin{figure} 
\begin{center}
\includegraphics{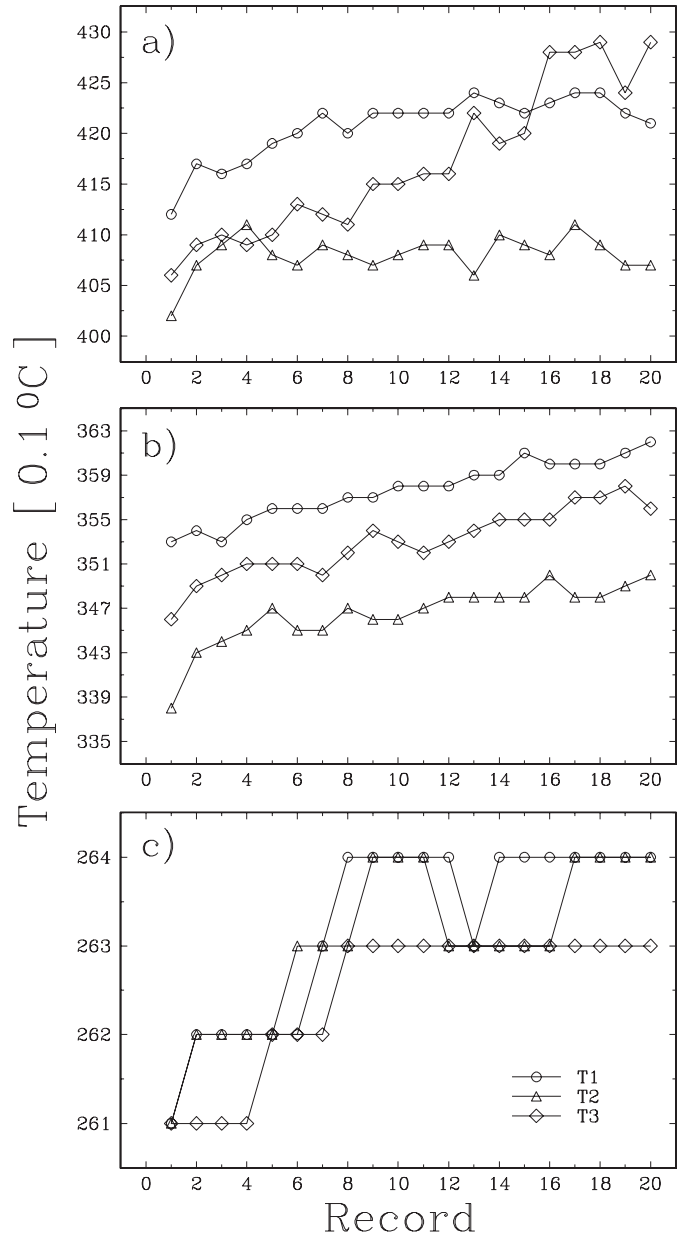}
\end{center}
\caption{Temperatures in positions 1, 2 and 3 for hard (a), soft (b) and normal (c) turbulence.}
\label{figura2}
\end{figure}

Figure {\ref{figura2}} shows the temperature for the three turbulence levels over 
the twenty records.  
Note that there are temperature gradients over the twenty records for the three 
turbulence levels and that the temperature ($T1$, $T2$ and $T3$, 
see Fig. \ref{figura1}) is not homogeneously distributed.  
So, an uniform flux of warm air and uniformity of turbulence along the path was 
not present. 

\section{Results and discussion}
\label{sec:results}

We transformed the twenty time records corresponding to the laser beam $x$ and $y$-coodinates ($M=5000$), under the three turbulence conditions, to the time-frequency domain by means of orthogonal discrete wavelet transform (ODWT)\cite{Daubechies92,Aldroubi96,Mallat99,Samar99} obtaining in this way the corresponding wavelet coefficients $C_j(k)$ series. In the present work, we consider eight wavelet resolution levels and cubic spline mother wavelet\cite{PRE3,Method01,Mairal02}. Note that the wavelet coefficients were non-overlapping for each scale. 

\begin{figure}
\begin{center}
\includegraphics{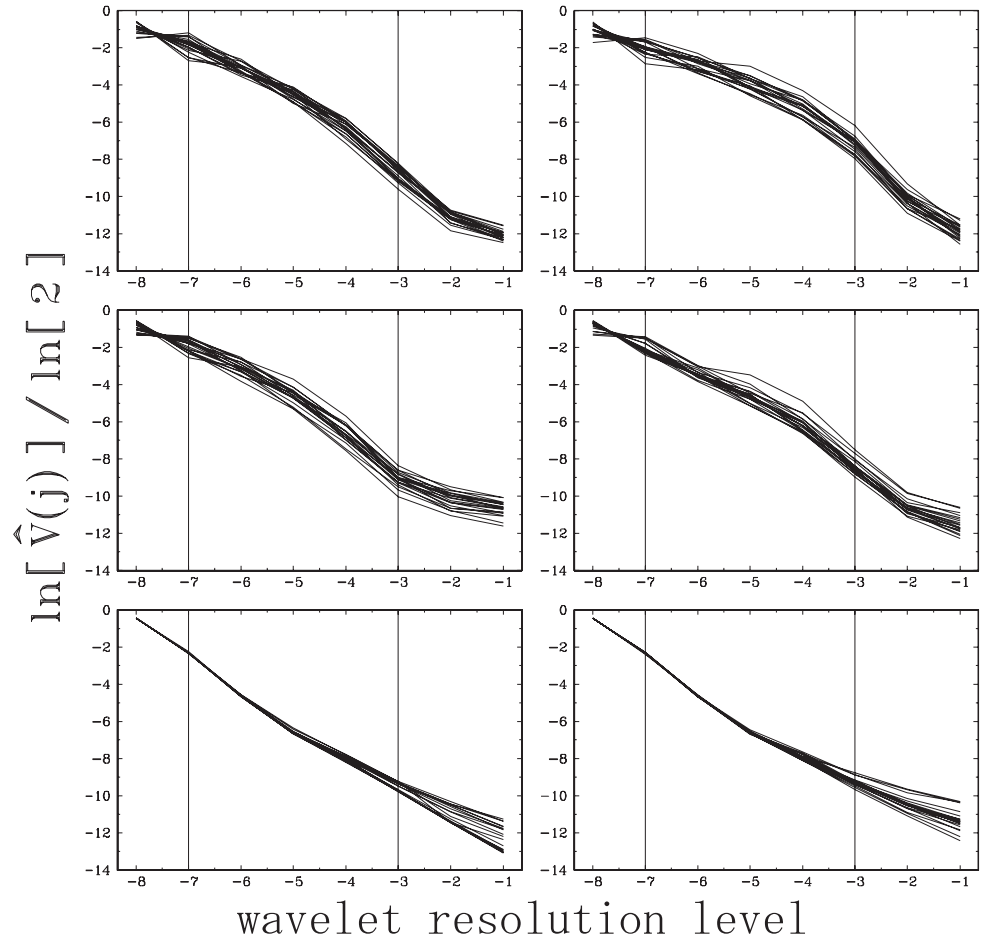}
\end{center}
\caption{
Evaluation of Hurst exponent for laser beam $x$-coordinate (left column) and 
$y$-coordinate, for hard (top), soft (center) and normal (bottom) turbulence. 
The vertical lines represent the scaling region used in the evaluation of $H$. 
}
\label{figura3}
\end{figure}

The wavelet coefficients were squared to obtain the associted wavelet energy and the estimator of wavelet coefficient variance $\hat{V}[j]$. To evaluate the Hurst exponent the procedure described at the end of Sec. {\ref{sec:fBmwavelet}} is used. Figure \ref{figura3} presents $\log_2\{{\hat V}[j]\}$ versus the resolution level $j$, for the $(x,y)$-coordinates and for the three turbulence conditions.  The vertical lines represent the scaling region used for the square fitting. The  values of the coefficient $H$ obtained were averaged over the twenty  records at each  turbulence condition for the $(x,y)$-coordinates respectively.  

On the other hand, each signal under analysis was divided in time windows of length $L=256$ and for each one the probability wavelet energy distribution was evaluated. With  these values the mean wavelet energy distribution and the mean NTWS, $\widetilde{S}_{WT}$, were obtained. This quantifier is taken as representative of the order-disorder of the whole signal. As before, the obtained  values $\widetilde{S}_{WT}$ were averaged over the twenty records at each turbulence condition for the $(x,y)$-coordinates respectively. 
  
\begin{figure} 
\begin{center}
\includegraphics[width=0.8\textwidth]{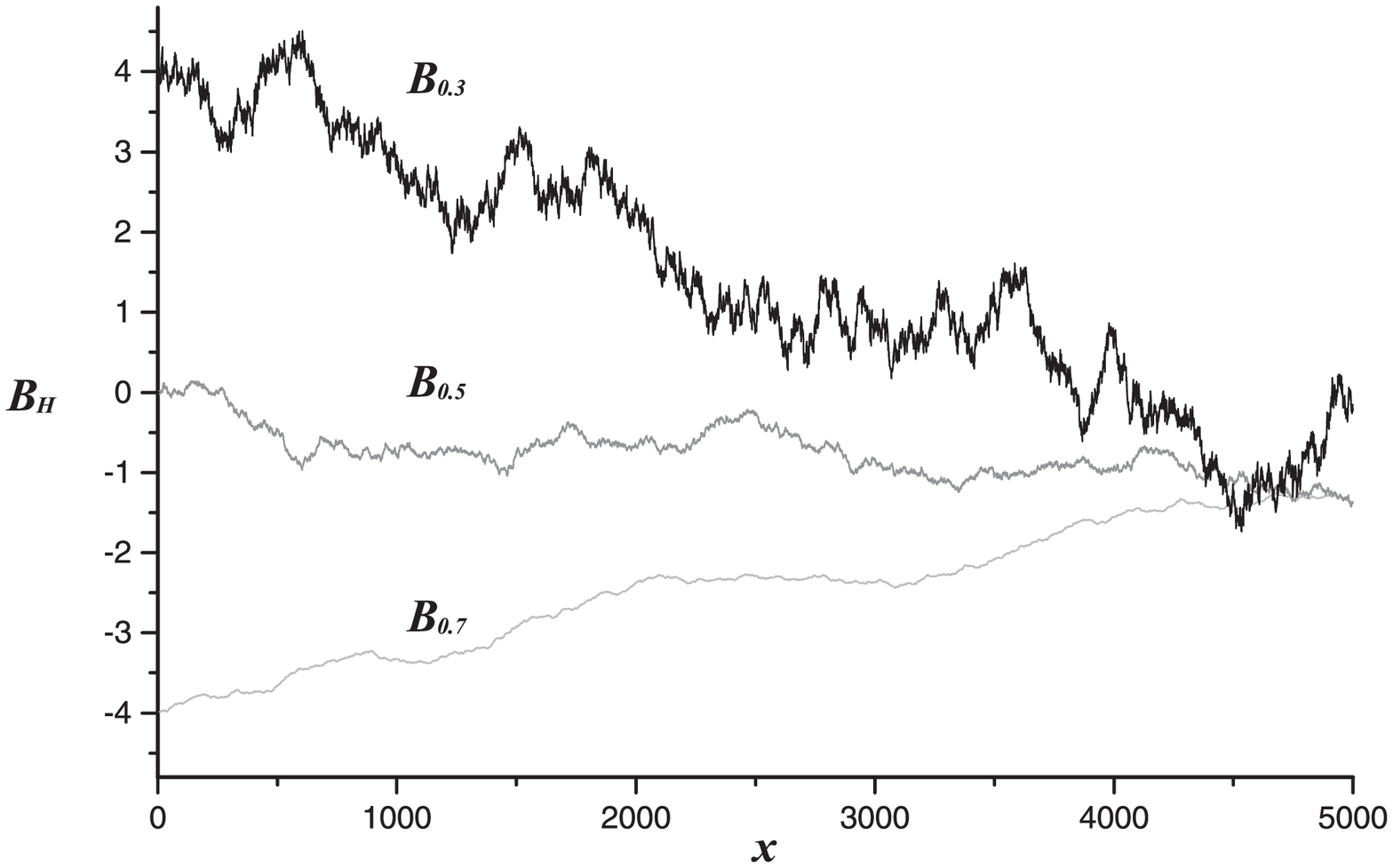}
\end{center}
\caption{Sample paths of fractional Brownian motions with $H=0.3$ (antipersistent), $H=0.5$ (standard brownian motion), and $H=0.7$ (persistent). These graphs were obtained by using the software \textbf{FRACLAB}\cite{url:frac}.}
\label{figura0}
\end{figure}

\begin{figure}
\begin{center}
\includegraphics{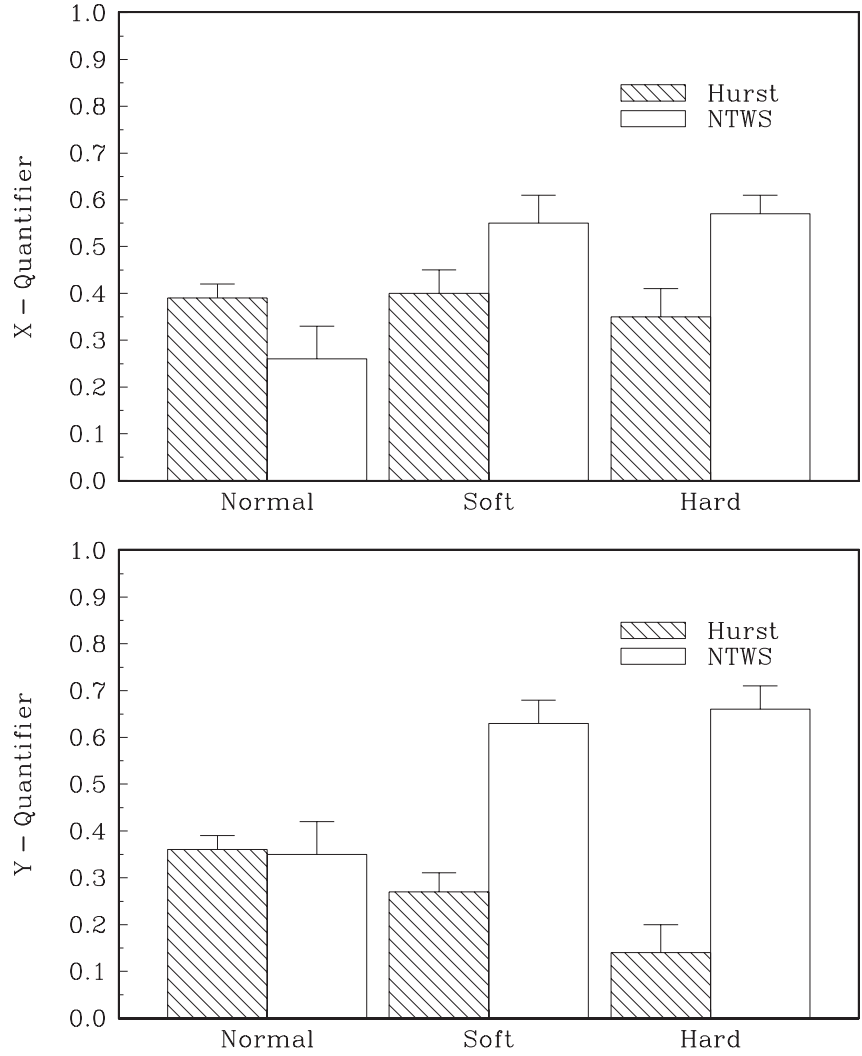}
\end{center}
\caption{Averages for Hurst exponents and mean NTWS (with their respective standard deviation errors) for the $x$ and $y$ coordinates in the case of normal, soft, and hard turbulence.}
\label{figura4}
\end{figure}

A comparison between average Hurst exponent and mean NTWS is given in Fig. {\ref{figura4}}. It is known the entropy is a measure of the order of a given system. In this case, the mean NTWS shows the same behavior for both coordinates. As the turbulence increases the mean NTWS does the same. But, the Hurst exponent discriminates between coordinates: for the $x$-axis no noticeable change is observed, and for the $y$-axis the Hurst exponent decreases (increasing the roughness, see Fig. \ref{figura0}) with the increasing turbulence. 

Thus, it is observed that the Hurst exponent is sensitive to the mean flow of the warm air. It distinguishes the anisotropy characteristic of the convective turbulence. Because the entropy measures the order obviously it does not detect the mean flow.

In any case, the value of the obtained quantities indicates that memory as well as self-similarity  and scale invariance  are  significant  property  of  these  time series.

In relation with the Hurst exponent a useful generalization  consists in allowing the singularity  exponent to become time-dependent $H(t)$, thus generating a new process $B_{H(t)}(x)$ such that
\begin{equation}
\mathbb{E}\left[(B_{H(t)}(t(x+\Delta x)-B_{H(t)}(t(x))^2\right]=\sigma^2\left|\Delta x\right|^{2H(t)}.
\label{eq-1c}
\end{equation}
In  such  a  situation  the increment  process  is  no  longer stationary. Then,  it is impossible to  globally  apply  the technique  mentioned  above  for estimating $H(t)$.   Provided  that variations of $H(t)$ are smooth enough,  the time series can be divided in time windows  where this requirement is satisfied and for each one the similar techniques based  on time-scale  energy  distributions  can be  applied locally\cite{peltier95}.  In any case, we must emphasize that for each one an scaling region must be defined. The time evolution of NTWS could be easily implemented\cite{Method01,Mairal02}. Moreover, the  NTWS  is  capable of detecting changes in  a  nonstationary signal due to the localization characteristic of the  wavelet transform, the computational time is  significantly shorter since the algorithm involves the use of wavelet transform in a multiresolution framework and, the NTWS is parameter-free and scaling region is not necessary for its evaluation. 

In  the  next future the intention of our project is  to  make experiments varying the intensity of turbulence by  adjusting the  voltage  applied to the heating element using  a  voltage controller, in  order  to  study  the  quantifiers   temporal evolution and characterize in a quantitative way the  dynamics of  the  process.  In  another  hand,  the same study will be  doing  in  outdoor experiments  at  different moments of the day to  characterize the ground atmospheric turbulence.

\newpage

\end{document}